\documentclass[prb,aps,emtex]{revtex4}
\usepackage{graphicx}
\usepackage{psfig}
\input epsf

\newcommand{\doublespace}{
    \renewcommand{\baselinestretch}{1.75}
    \large\normalsize}
\newcommand{\be}{\begin{equation}}
\newcommand{\ee}{\end{equation}}
\newcommand{\bea}{\begin{eqnarray}}
\newcommand{\eea}{\end{eqnarray}}

\newcommand{\erfc}{\text{erfc}}
\begin{document}
\doublespace

\title{Different regimes of F\" orster energy transfer between an
    epitaxial quantum well and a proximal monolayer of semiconductor
    nanocrystals }
\author{\v S. Kos$^{(1)}$, M. Achermann$^{(2)}$, V.I. Klimov$^{(2)}$,
    D.L. Smith$^{(1)}$}
\address{{\rm (1)}Theoretical Division, Los Alamos National
    Laboratory, Los Alamos, New Mexico 87545}
\address{{\rm (2)}Chemistry Division,
    Los Alamos National Laboratory, Los Alamos, New Mexico 87545}
\begin{abstract}
    We calculate the rate of non-radiative,
    F\"orster-type energy transfer (ET) from an excited epitaxial
    quantum well (QW) to a proximal monolayer of
semiconductor nanocrystal quantum dots (QDs).
    Different electron-hole configurations in the QW
    are considered as a function of temperature and
excited electron-hole density.
    A comparison of the theoretically determined ET rate and
    QW radiative recombination rate shows that,
depending on the specific conditions,
   the ET rate is comparable to or even greater than
the radiative recombination rate.  Such efficient
F\"orster ET
    is promising for the implementation of ET-pumped,
nanocrystal QD-based light emitting devices.
\end{abstract}

\maketitle

\section{Introduction}
Modern colloidal chemistry allows the fabrication of semiconductor
nanocrystal quantum dots (QDs)
with nearly atomic precision in a wide range of sizes and
shapes.\cite{Alivisatos,murray:93:1,Hines,Dabbousi,peng:2001:1,VictorBook}
Nanocrystal QDs exhibit high photoluminescence
(PL) quantum yields \cite{Hines,Dabbousi,peng:2001:1}
   and size-controlled emission spectra.
Nanocrystals can also be easily manipulated into various two-dimensional (2D)
and 3D assemblies.\cite{murray:95:1,achermann:JPCB:2003,Crooker}
All of these properties make nanocrystal QDs attractive building blocks for
   various optical devices including color-selectable light-emitters.
A major problem associated with the realization of nanocrystal
QD-based light emitters is that the
electrical injection of carriers into nanocrystals is complicated by
the presence of an insulating passivation layer.
All previous attempts to electrically contact nanocrystals have
utilized hybrid
inorganic/organic composites comprising
conducting polymers.\cite{Colvin,dabbousi:95:1,banin:2002:LED,Coe}
However, the performance of these devices is severely limited by
low carrier mobilities in both nanocrystal and polymer components and
poor polymer stability with respect to photooxidation.

Recently, we have presented an alternative,
``noncontact" approach to injecting carriers into
nanocrystal QDs via nonradiative
energy transfer (ET) from a proximal epitaxial quantum well (QW).\cite{Marc}
The experiments revealed efficient energy outflow from the QW,
which was accompanied by a complementary energy inflow into a
dense monolayer of nanocrystals assembled on top of the QW.
The measured ET rates were very fast allowing for the efficient
pumping of nanocrystal QDs.

In this paper we develop the theoretical framework to model
F\"orster-type ET from an epitaxial QW to a monolayer of nanocrystal 
QDs.
Our approach is conceptually similar to the theory pioneered by Agranovich
and co-workers \cite{agranovich:1998:1,Basko} who studied ET from a QW
to an adjacent, infinitely thick layer  of organic molecules. In 
addition to considering a different type of acceptors (semiconductor 
nanocrystals vs. organic molecules), and a different ``geometry" of 
the ET system (ET to a single proximal monolayer vs. ET to an 
infinitely thick layer), in the present work we calculate both ET and 
radiative decay rates, which allows us to analyze the ET efficiency 
as a function of temperature and excitation density. We consider the 
situations for which the electronic excitations in the QW are present 
either in the form of free electrons and holes or Coulombically bound 
electron-hole pairs (excitons). We also account for the effects of 
exciton localization at defect states. Furthermore, we consider two 
types of resonant QD acceptor states that can be treated either as a 
dense quasi-continuum (applicable to high-energy QD states located 
well above the band edge) or the narrow, atomic-like resonances 
(applicable to near-band-edge QD states).   Finally, we apply the 
developed theory to model our experiments on energy transfer between 
the InGaN QW and a proximal monolayer of CdSe nanocrystal QDs. We 
find a remarkable agreement between our experimental observations and 
the results of the calculations performed for the case of free 
electrons and holes in the QW. Independent studies of 
pump-intensity-dependent QW photoluminescence (PL) confirm that under 
our experimental conditions the QW excitations can indeed be well 
described in terms of unbound electrons and holes.

The paper is organized as follows: We introduce the general formalism
in Section \ref{General formulas}.
Then, in Section \ref{Excitons in the QW} and \ref{Free carriers}, we study
coupling of QW excitation to
   high-energy states of the QDs that form a dense, quasi-continuous
   spectrum. 
  In Section \ref{Excitons in the QW} we investigate the excitation
density and temperature regime, in which the excited
electron-hole pairs in the QW
are bound
into non-interacting excitons described by classical statistics. At high
temperatures within this regime, the excitons are mobile in the QW
and the F\"orster rate dominates the radiative decay rate. 
At lower temperatures, the localization of excitons at defects
decreases the efficiency of F\"orster ET. 
In Section \ref{Free carriers},
   we study the density and temperature regime, in which the
carriers form a two-dimensional (2D) plasma in the QW.
As the carrier density increases for a
given temperature, the plasma experiences a transition from the 
non-degenerate to the degenerate regime. In this  case, we find 
that the F\"orster rate is always greater than the radiative rate and 
both rates reach maximum around the degeneracy
   temperature for the holes.

  In Section \ref{Discrete QD states}, we examine the case in which 
QW excitations couple to discrete, low-lying QD states with linewidths
   that are smaller than the characteristic energy of the motion of charge
   carriers in the QW. We find that the ET rate decays exponentially with
   increasing  QW-QD distance in contrast to the power law found for the
   situation in which the ET occurs into the quasi-continuum of 
high-energy QD states.

Finally, in Section \ref{Experiment}, we describe experimental results 
obtained for the ET structure composed of an InGaN QW
   and CdSe QDs. The analysis of these results indicates that the 
non-degenerate free carrier case best describes 
   the structure studied experimentally. We compare the ET rates, 
radiative decay rates, and ET efficiencies obtained from the theory
    with those measured experimentally and find good agreement.

  A summary of our results is presented in Section \ref{Summary and 
conclusion}.

\begin{figure}
\begin{minipage}[t]{3in}
\epsfxsize = 3in
\epsfbox{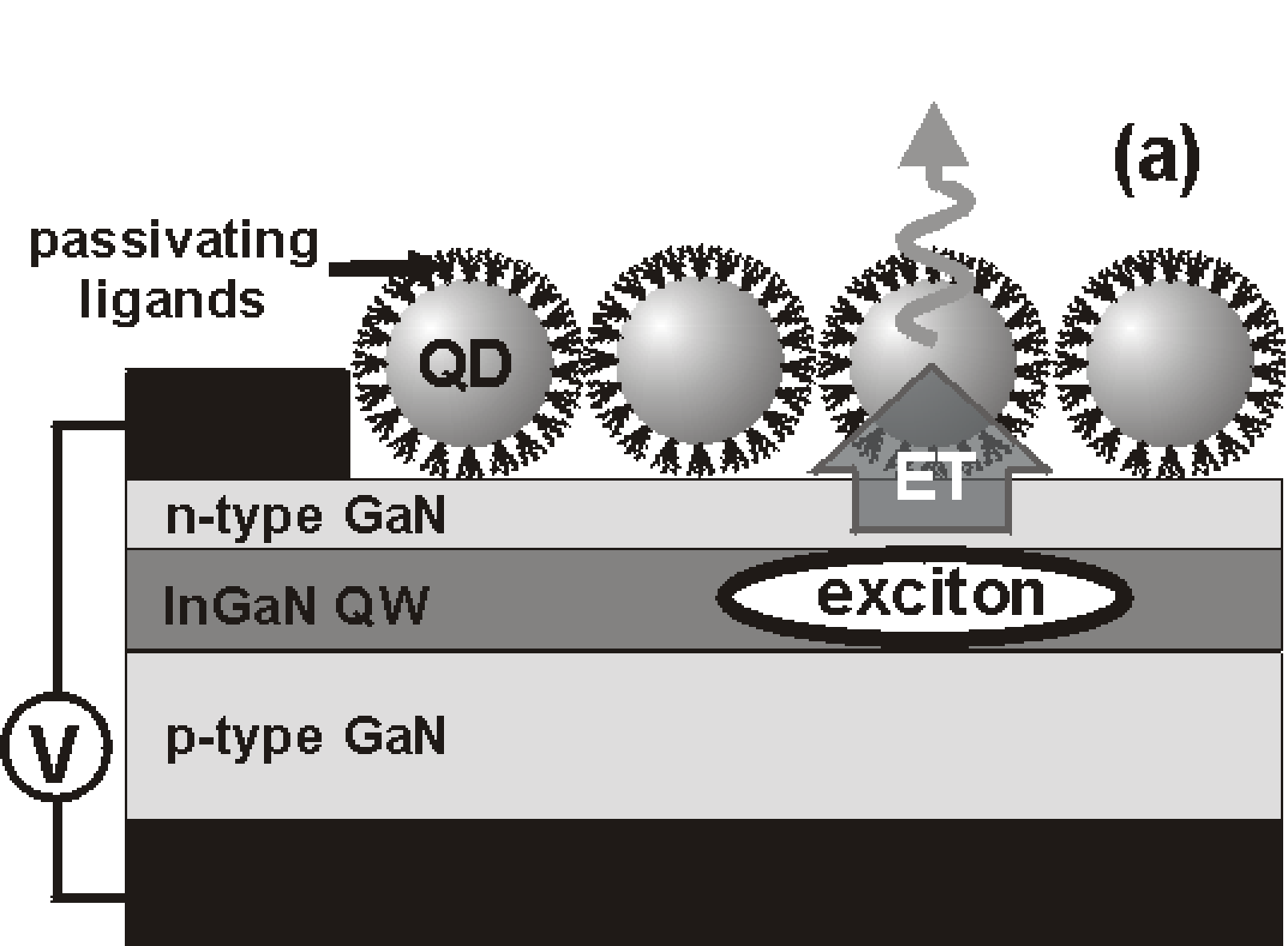}
\end{minipage}\
\begin{minipage}[t]{3in}
\epsfxsize = 2.4in
\epsfbox{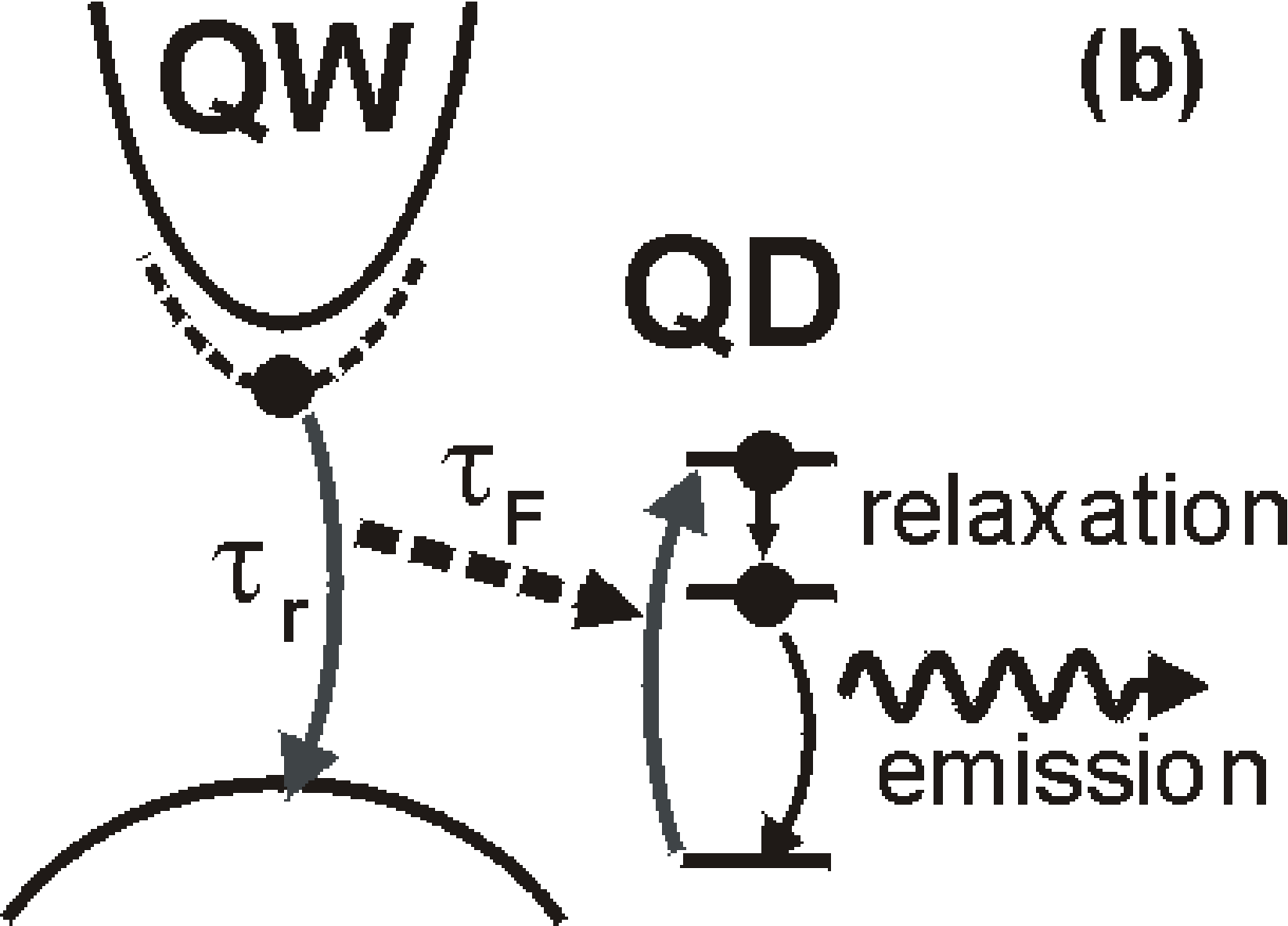}
\end{minipage}

\caption{(a) An example of a hybrid
QW/nanocrystal-QD structure that can be used
for ``non-contact" pumping of nanocrystals via
nonradiative ET. The structure consists of an
InGaN QW sandwiched between GaN
barriers. A monolayer of CdSe/ZnS core/shell
nanocrystals capped with organic molecules is
assembled on the surface of the thin
top barrier. The structure is driven electrically
using metal contacts attached to the barrier
layers. (b) Carrier relaxation
and ET processes in the hybrid QW/nanocrystal
structure. The QW-to-QD ET competes with
recombination processes in the QW.
High-energy excitations created in the
nanocrystals through ET rapidly relax to the
nanocrystal band edge, which prevents
backtransfer. The relaxed excitations recombine,
producing emission with the color determined by
the QD size.}
\label{fig with geometry}
\end{figure}



\section{General framework}
\label{General formulas}

The geometry of the studied structure is shown in
Fig. \ref{fig with geometry}. We consider an
epitaxial semiconductor heterostructure, where a
QW
has been grown between a thick bottom and a thin top barrier layer.
On top of the QW structure a monolayer of nanocrystal QDs has been assembled.
The derived formulas are not material specific; the numerical values
of the ET rates are computed in the subsequent sections for a
combination of an InGaN QW
and CdSe nanocrystal QDs. The distance $R$ between the
centers of the QD and the QW is determined by the QW and the top barrier
widths, the length of the ligand molecules, which surround the QD, and
the QD size. We shall simplify the calculations by assuming that the size of
the QDs and the width of the QW are small compared to $R$. In the
experiments described in Ref.\ 14, these quantities are
of the same order of magnitude as $R$. However,
taking into account the actual dimensions
of the QDs and the QW changes the results by a few percent only (as verified by
exact calculations assuming a non-zero width of the QW).

The F\"orster process transfers an electron-hole excitation from the
QW to the QD via the electrostatic interaction
\be
\label{electrostatic interaction}
H_I^F={e^2\over \epsilon}\int d^3 r_D d^3 r_W \sum\limits_{\alpha,\beta}
{\psi^{\dag}_{\alpha}({\bf r}_D)\psi_{\alpha}({\bf r}_D)
\psi^{\dag}_{\beta}({\bf r}_W)\psi_{\beta}({\bf r}_W)
\over
|{\bf r}_D-{\bf r}_W|},
\ee
where $\epsilon $ is the average of the high
frequency dielectric constants of the
QW and air.
The initial and final states participating in ET are
\bea
|i\rangle & = & |ex _W\rangle |GS _D \rangle
\eea
and
\bea
|f\rangle & = & |GS _W\rangle |ex _D \rangle,
\eea
where $|GS \rangle$ and $|ex \rangle$ are the ground state and the
excitated state, respectively (subscripts ``W'' and ``D'' denote QW
and QD, respectively).
The matrix elements of the density operators in the
numerator of Eq. (\ref{electrostatic interaction}) are zero for our
initial and final states, therefore, we use
the dipolar expansion of the matrix element. The
conduction and  valence bands have $s$ and $p$ symmetry,
respectively, so the dipolar matrix elements are non-zero. Thus, we
can use the dipolar approximation for the transition matrix element
for a single dot
\be
\langle f|H_I^F |i\rangle = {e^2 \over \epsilon}
\int d^2 r_W
{{\bf d}_D^* ({\bf r}_D)\cdot {\bf d}_W ({\bf r}_W)
-
3\left({\bf d}_D^* ({\bf r}_D) \cdot
{{\bf r}_D-{\bf r}_W \over |{\bf r}_D-{\bf r}_W|}\right)
\left({\bf d}_W ({\bf r}_W) \cdot
{{\bf r}_D-{\bf r}_W \over |{\bf r}_D-{\bf r}_W|}\right)
\over
|{\bf r}_D-{\bf r}_W|^3},
\ee
with
\be
{\bf d} ({\bf r})
=\Psi ({\bf r},{\bf r})
\langle u_v |{\bf r}|u_c \rangle,
\ee
where $u_c$ and $u_v$ are the periodic functions that enter the Bloch
wave functions for the
conduction and valence bands, respectively, and the envelope wave
function $\Psi $ is assumed to vary on a much larger length scale than the
lattice constant. To calculate the transition rate from the Fermi 
Golden Rule, we
square the modulus of this matrix element, multiply by the
energy-conserving delta
function, sum over the final and initial states
(weighted with
a thermal distribution function), and finally multiply by the number of the
dots.

We calculate the radiative transition rate in a QW by using the Fermi Golden
Rule with the interaction Hamiltonian
\be
H_I^r=-{1\over c}\int d^3r {\bf j}({\bf r})\cdot {\bf A}({\bf r}).
\ee
The transition matrix element now is
\be
\label{coupling to elmg field}
\langle {\bf k},\lambda |H_I^r|ex _W\rangle
\equiv
-{1\over c}\int d^3r
\langle GS_W |{\bf j}({\bf r}) |ex _W\rangle \cdot
\langle {\bf k},\lambda |{\bf A}({\bf r})|0\rangle ,
\ee
where $|{\bf k},\lambda \rangle$ is a one-photon state with
wave vector ${\bf k}$ and polarization $\lambda $, and $|0 \rangle$
is the photon vacuum. Here, we consider only spontaneous
emission; stimulated emission may be important for  free carriers
below the degeneracy temperature discussed in Section
\ref{Free carriers}. Using the
canonical commutation relations between position and momentum
operators, we obtain
\be
\label{general current matrix element}
\langle GS_W |{\bf j}({\bf r}) |ex _W\rangle
=
-i E_G e {\bf d}_W({\bf r}),
\ee
where we have approximated the exciton energy by the band-gap energy
$E_G$. Again, to determine the radiative transition rate, we need to multiply the square of the modulus of
this transition matrix
element by the energy-conserving delta function, and sum over all the
one-photon states.

\section{Excitons in the QW}
\label{Excitons in the QW}

The conduction band is two-fold degenerate as a result of the two spin
projections; the orbital part of both bands is the same, and has 
$s$ symmetry. The valence band is four-fold degenerate in the center
of the Brillouin zone, but the spatial confinement in the direction
perpendicular to the QW splits the degeneracy. We
will only take into account the low-energy,
heavy-hole band.

First we consider the case of a free exciton. For the envelope function, we
make a separation of variables into the center-of-mass motion described
by a plane wave and the relative motion assuming the exciton in a hydrogen
1s state with Bohr radius $a_B$. This results in a
dipolar matrix element

\be
{\bf d_W} ({\bf r})
=\Psi ({\bf r},{\bf r})
\langle u_v |{\bf r}|u_c \rangle=
{e^{i{\bf K} \cdot {\bf r}}\over \sqrt{{\pi\over 2}A a_B^2}}{d_W 
\over \sqrt{2}}(\hat x \pm i\hat y),
\ee
in which $A$ is the QW area, $d_W$ is the QW dipole moment and ${\bf K}$ 
is the center-of-mass momentum.

In this case, the matrix element of the F\"orster transition is
\be
\label{itinerant transition matrix element}
\langle f|H_I^F |i\rangle =- {e^2 \over \epsilon}
\sqrt{2\over \pi A a_B^2}
({\bf d}_D^* \cdot \nabla )  d_W( \partial _x \pm i \partial _y)
\phi _{\bf K}({\bf R}),
\ee
with
\be
\phi _{\bf K}({\bf R}) \equiv
\int d^2 r {e^{i{\bf K}\cdot {\bf r}}\over
|{\bf r}-{\bf R}|}
=2\pi {e^{i{\bf K}\cdot {\bf \rho}-Kz} \over K},
\ee
where ${\bf \rho }$ and $z$ are the in-plane and out-of-plane
components of the vector ${\bf R}$, respectively.
The dipolar matrix element of the dot, ${\bf d}_D$, has a particular
value and a random orientation.
Squaring the absolute value of the F\"orster  matrix element, 
averaging over the
random direction of ${\bf d}_D$, summing over the high-energy QD
excitons with a smoothly varying density of states $N_D(E)$, and summing
over the QDs, we obtain the F\"orster transfer rate (the inverse of the
ET time, $\tau_F$):
\be
\label{1 over tau F (K)}
{1\over \tau_F(K)}={8\pi \over 3}
\left({e^2 \over \epsilon}\right)^2
|d_D|^2 |d_W|^2 N_D(E_G)
{n_D \over a_B^2} K^2 e^{-2KR},
\ee
where $n_D$ is the areal density of the QDs and $R$ is the QW-QD 
separation in the $z$-direction.

To model excitons in a real QW, we have to consider width 
fluctuations, alloy disorder or impurities that
can localize the exciton. If the length scale of such a trap is much 
larger than
$a_B$, the relative motion of the exciton will remain unchanged, but the
center-of-mass wave function will now be localized instead of being a
plane wave. We assume that all the traps in the QW localize the
excitons into states with a characteristic binding energy $E_T$ and
localization length $\xi $. An analytical evaluation of the matrix elements
is possible only for special exciton envelope functions, such as a 
modified Lorentzian,
\be
\Psi _{\xi }(r) =\sqrt{2\over \pi}
{\xi ^{-1}\over \left[1+\left({r\over \xi}\right)^2\right]^{3/2}},
\ee

leading to an ET rate of the bound excitons of
\be
\label{1 over tau F loc}
{1\over \tau _{F,loc}} = 4\pi
\left({e^2 \over \epsilon}\right)^2
|d_D|^2 |d_W|^2 N_D(E_G) n_D \left( {\xi \over a_B}\right)^2
{1\over (\xi +R)^4}.
\ee
For a different center-of-mass wave function, the
functional dependence will be somewhat different, but the asymptotic
behavior in the two limits $\xi >> R$ and $\xi << R$ will be the same,
hence this is a suitable interpolation between the two limits.

We obtain the total F\"orster transfer rate by
averaging (\ref{1 over tau F (K)})
and (\ref{1 over tau F loc}) using the Boltzmann
distribution
\bea
\label{combined Forster rate for excitons}
{1\over \tau _F} & = &
{n_T e^{-E_T\over T}{1\over \tau _{F,loc}}
+\int {d^2 K\over (2\pi)^2} e^{-K^2 \over 2MT} {1\over \tau _F(K)}
\over
n_T e^{-E_T\over T}
+\int {d^2 K\over (2\pi)^2} e^{-K^2 \over 2MT}}
\nonumber \\
& = &
4\pi \left({e^2 \over \epsilon}\right)^2
|d_D|^2 |d_W|^2 N_D(E_G) {n_D \over a_B^2}
{n_T e^{-E_T\over T}{\xi ^2 \over (\xi + R)^4}
+ {MT\over 2\pi } {4\over 3R^2}f(2MTR^2)
\over
n_T e^{-E_T\over T}+{MT\over 2\pi}},
\eea
where $n_T$ is the areal density of the traps,
$M\equiv m_e+m_h$ is the mass of the exciton,
and
\be
\label{definition of f}
f(x)\equiv {1\over x}\int\limits _0 ^{\infty}
d\kappa \kappa ^3 e^{-2\kappa-{\kappa ^2 \over x}}
={x\over 4}[2(1+x)-
e^x\sqrt{\pi x}(3+2x)\erfc(\sqrt{x})]
.
\ee

The performance of a device that relies on ET-pumping of colloidal QDs
from a QW is determined by the ET efficiency ($\eta$), which is a function 
of both the ET rate and the total rate of all recombination processes in the QW ($\tau _r^{-1}$):
$\eta = \tau _F^{-1}/(\tau _F^{-1}+\tau _r^{-1})$. 
An ultimate limit on the lifetime of QW excitations is imposed by the radiative decay,
which determines an upper limit of the ET efficiency.
We calculate the radiative recombination rate from
(\ref{coupling to elmg field}). The matrix element of the one-photon
transition is
\be
\langle {\bf k},\lambda |{\bf A}({\bf r})|0\rangle
=
\sqrt{2\pi c\over k} {\bf e}_{\lambda }^*({\bf k})
e^{-i{\bf k}\cdot {\bf r}}.
\ee
The integral in (\ref{coupling to elmg field}) will set the
in-plane component of
${\bf k}$ equal to ${\bf K}$ and give the radiative rate of a free exciton
\cite{Runge}

\be
\label{1 over tau r for a single itinerant exciton}
{1\over \tau _{r}(K)}=
{1\over \pi}(e d_w)^2 \left({E_G \over c}\right)^2{1\over a_B^2}
{1\over \sqrt{\left({E_G\over c}\right)^2-K^2}}
\left(2-{c^2K^2\over E_G^2}\right) \theta (E_G-cK).
\ee
For a localized exciton, we find
\be
{1\over \tau _{r,loc}}={32\over 3\pi}
(e d_W)^2 \left({E_G \over c}\right)^3 {\xi ^2\over a_B^2},
\ee
resulting in the total recombination rate (obtained by
averaging with respect to the Boltzmann
distribution)
\be
\label{combined recombination rate for excitons}
{1\over \tau _r} = {4\over 3\pi}
(e d_W)^2 \left({E_G \over c}\right)^3 {1\over a_B^2}
{{1\over 2\pi}+n_T e^{- E_T \over T}8\xi^2
\over
{MT\over 2\pi}+n_T e^{- E_T \over T} }.
\ee

\begin{figure}
\epsfxsize=7 in
\epsfbox{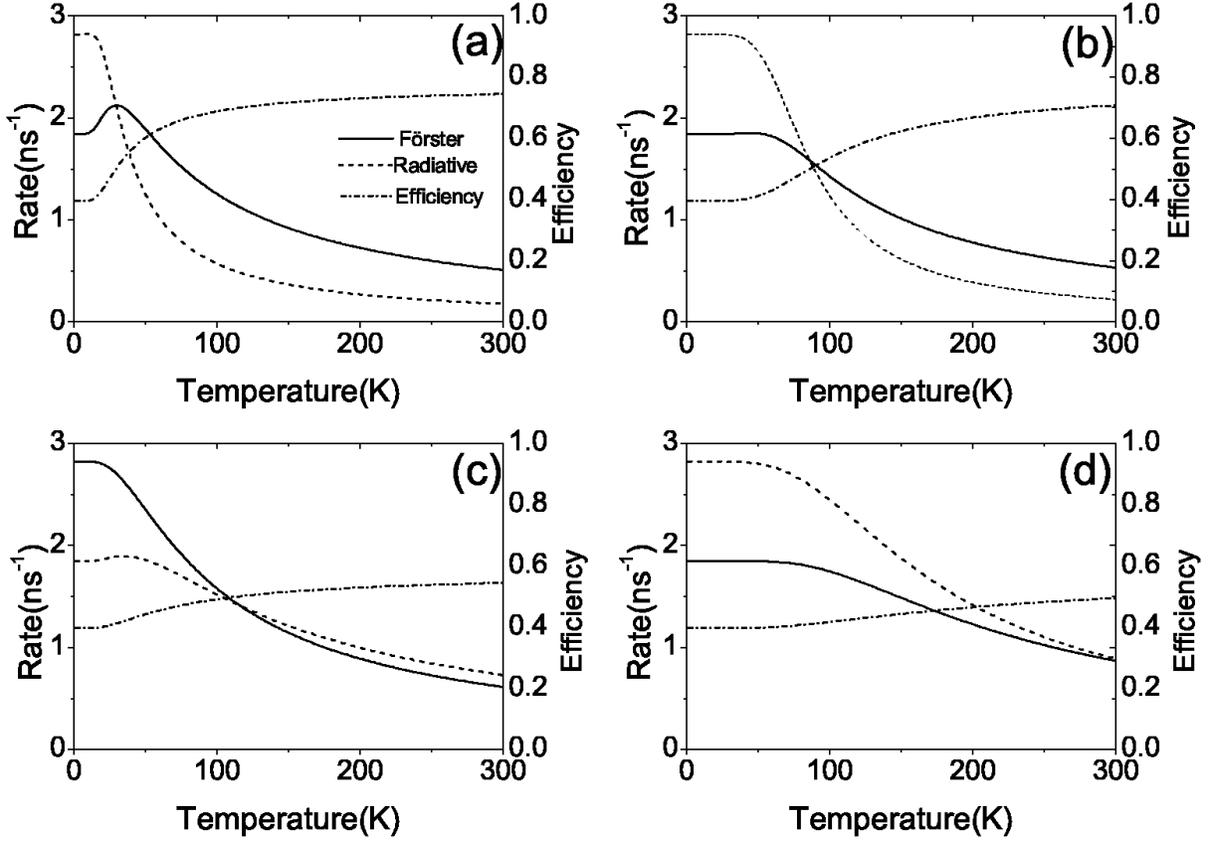}
\caption{The F\"orster (solid line) and radiative (dashed line)
transition rates calculated for QW excitons for $\xi=30$ \AA\ and 
(a) $n_T=10^{11}\text{ cm}^{-2}$ and $E_T=-0.005$ eV,
(b) $n_T=10^{11}\text{ cm}^{-2}$ and $E_T=-0.02$ eV,
(c) $n_T=10^{12}\text{ cm}^{-2}$ and $E_T=-0.005$ eV,
(d) $n_T=10^{12}\text{ cm}^{-2}$ and $E_T=-0.02$ eV. The dashed-dotted line
is the efficiency of the F\"orster transfer.}
\label{Fig2}
\end{figure}

In Fig. \ref{Fig2}, we plot
the F\" orster (\ref{combined Forster rate for excitons})
and the radiative (\ref{combined recombination rate for excitons})
transition rates, along with the efficiency of the F\"orster transfer
as a function of
temperature. We use the
following physical parameters from the experiment in Ref.\ 14,
where a monolayer of CdSe nanocrystal QDs with a 
$19 \text{ \AA}$ radius
has been deposited on an InGaN QW with a $30\text{ \AA}$ width:
$d_D=5.2 $ \AA \cite{Crooker}, $d_W=2.9$ \AA \cite{Lawaetz},
$N_D(E_G)=17.3 $ eV$^{-1}$
(determined from the QD absorption spectra), $E_G=3.1$ eV,
$n_D=2\times 10^{12} \text{ cm}^{-2}$, $a_B=27.8 $ \AA,
$\epsilon =3.6$ [the average of the
high frequency dielectric constant of GaN
($\epsilon_{W,\infty}=6.2$) and air ($\epsilon_{air}=1$)],
$M=m_e + m_h=0.2m_0+0.8 m_0$, $R=81 $ \AA. In addition, we assume
$\xi = 30 $ \AA{} (the lower bound for $\xi $ is
$a_B$), $n_T=10^{11}$ cm$^{-2}$, and $E_T=-0.005$ eV in (a) and
$E_T=-0.02$ eV in (b). For the static dielectric constant of GaN
$\epsilon _{W,0}=8.9$, we obtain the exciton binding energy of 0.0528 eV,
which corresponds to 675 K. We plot the radiative and ET rates and the ET efficiency up to
 room temperature, which is roughly half of the latter value.

At high temperatures
$T>>1/2MR^2 \simeq 5 K$, both $1/\tau _F$ and $1/\tau _r$ behave as
$1/T$ with a bigger prefactor for $1/\tau _F$. As the temperature
decreases, both rates increase, but the increase of $1/\tau _F$
becomes slower than $1/T$, and eventually the two rates cross and
$1/\tau _F$ becomes smaller than $1/\tau _r$. The ET efficiency,
therefore, increases as the temperature increases up to the crossing point and
levels off at temperatures above this point; the individual rates, however, 
decrease as the temperature is increased further. Thus, the optimum
temperature for the ET is around the crossing point
between $1/\tau _F$ and $1/\tau _r$, which occurs at temperature on the
order of $1/2MR^2$ if the effect of the exciton localization is
negligible. The crossing point shifts to higher
temperatures and the ET efficiency is decreased in the case of stronger
exciton localization. Localized excitons dominate
the rates in the limit $T\rightarrow 0$, thus making the rates at the crossing
point smaller. Indeed, $\tau _{r,loc}^{-1}$ grows relative to
$\tau _{F,loc}^{-1}$ with increasing localization length $\xi$, and
the importance of the
localized term grows both with $n_T$ [compare Fig.\ref{Fig2} (a)/(b)
with (c)/(d)] and with $|E_T|$ [compare Fig.\ref{Fig2} (a)/(c)
with (b)/(d)]. We note that the order of magnitude of the transition
rates does not change as we vary the localization
parameters (compare different panels in Fig. 2).


\section{Free carriers in the QW}
\label{Free carriers}

When the density of electrons or holes $n_{eh}$
\cite{footnote1}
exceeds $1/a_B^2$ (so that the kinetic energy dominates the Coulomb
interaction), or the temperature is higher than the exciton binding
energy, then the charge carriers do not form excitons but rather
stay unbound. We assume that the energies of the electrons and
holes are much larger than $E_T$, so that we can treat the charge
carriers as free particles described by plane waves
\be
\Psi ({\bf r},{\bf r}) =
{e^{i{\bf K} \cdot {\bf r}}\over A }.
\ee
The use of these wavefunctions modifies the normalization in
(\ref{1 over tau F (K)}). In the classical (non-degenerate) case, 
that is $n_{eh}<2m_eT/2\pi, 2m_hT/2\pi$, we can express the ET rate
(after averaging over the thermal distribution of free
carriers) as

\be
\label{Forster classical rate}
{1\over \tau _{F,class}}
={8\pi^2 \over 3}
\left({e^2 \over \epsilon}\right)^2
|d_D|^2 |d_W|^2 N_D(E_G) n_D n_{eh} {1\over R^2} f(2MTR^2).
\ee
The latter expression indicates that the ET rate in the free-carrier
case is proportional to $n_{eh}$.

For $n_{eh}>2m_eT/2\pi,2m_hT/2\pi$, both the electrons and the holes form
degenerate Fermi gases and the total F\"orster rate will be
\be
{1\over \tau _{F,deg}}
=
\sum\limits _{{\bf k}_e,{\bf k}_h}^{\sqrt{2\pi n_{eh}}}
{1\over \tau _F(K)}.
\ee
We convert the sum over momentum into an integral that can be
calculated explicitly, holding ${\bf k}_e+{\bf k}_h={\bf K}$.
The F\"orster rate per e-h pair in the
degenerate regime is
\be
\label{Forster degenerate rate}
{1\over \tau _{F,deg}}
= {2\over 3}
\left({e^2 \over \epsilon}\right)^2
|d_D|^2 |d_W|^2 N_D(E_G) n_D {1\over R^4} g(2\sqrt{2\pi n_{eh}R^2}),
\ee
where
\be
g(x)\equiv x^4 \int\limits_0^1 d\kappa
\left(
{\pi \over 2}- \arcsin\kappa
-\kappa \sqrt{1-\kappa ^2}\right)
\kappa ^3 e^{-2\kappa x}.
\ee

The corresponding recombination rates will be obtained from the decay rate
calculated for a single e-h pair with the center-of-mass momentum 
${\bf K}$, which
is the same as
(\ref{1 over tau r for a single itinerant exciton}) except that
$a_B^2$ in the denominator is replaced with the total QW
area $A$.

For the classical case ($n_{eh}<2m_eT/2\pi,2m_hT/2\pi$) we obtain

\be
\label{radiative classical rate}
{1\over \tau _{r,class}} = {2\over 3}
(e d_w)^2 \left({E_G \over c}\right)^3
{n_{eh} \over MT},
\ee
whereas for the degenerate case, $n_{eh}>2m_eT/2\pi,2m_hT/2\pi$,
the radiative recombination rate is
\be
\label{radiative degenerate rate}
{1\over \tau _{r,deg}}
\simeq
\sum\limits _{{\bf k}_e,{\bf k}_h}^{\sqrt{2\pi n_{eh}}}
{1\over \tau _{r,free}(K)}= {1\over 6\pi}
(e d_w)^2 \left({E_G \over c}\right)^3.
\ee

Finally, if $m_h>>m_e$, we  have an intermediate regime
\be
\label{mixed condition}
2m_eT/2\pi<n_{eh}<2m_hT/2\pi,
\ee
that is, the electrons are degenerate and the holes are classical. We 
can obtain the F\"orster and
radiative rates from formulas
(\ref{Forster classical rate})
and
(\ref{radiative classical rate}), respectively,
if we replace $M$ with
\be
m_e{E_{Fe}\over T}+m_h,
\ee
where  $E_{Fe}= 2\pi n_{eh}/2m_e$.
Since
\be
m_e{E_{Fe}\over T}=m_h{2\pi n_{eh}\over 2m_hT}<m_h
\ee
in the intermediate regime, the F\"orster and
radiative rates behave as in the classical regime with the total mass
replaced by the mass of the hole.

\begin{figure}
\epsfxsize = 7 in
\epsfbox{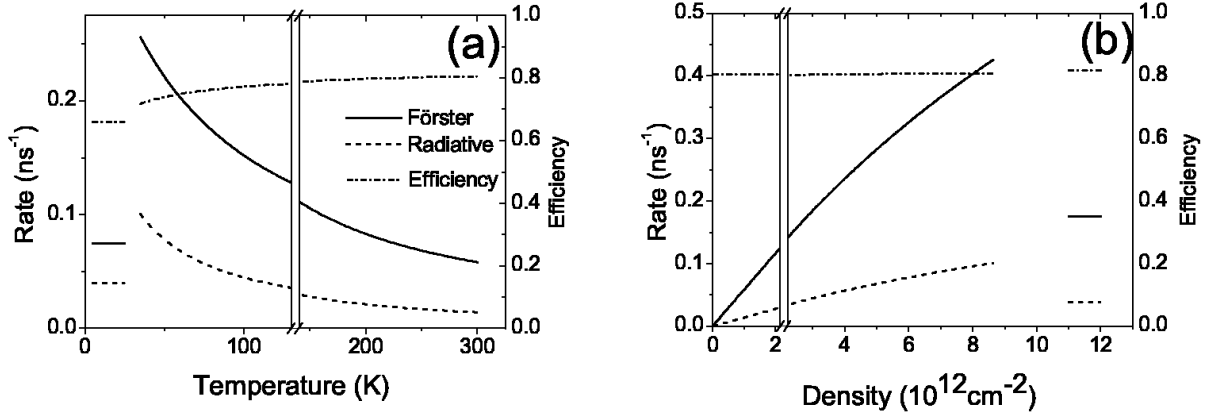}
\caption{The F\"orster (solid line) and radiative (dashed line)
    transition rates for QW classical free carriers, along with the ET efficiency (dashed-dotted line). 
    (a) The calculations are shown for
$n_{eh}=10^{12}\text{ cm}^{-2}$ as a function of $T$ above the
degeneracy temperature 34.7 K. (b)
The calculations are shown for $T=300$ K as a function of $n_{eh}$ below the degeneracy density
$8.6\times 10^{12} \text{ cm}^{-2}$. The straight-line segments  indicate the
asymptotic degenerate values in the limit $T\rightarrow 0$ for the
F\"orster (0.0749 ns$^{-1}$) and radiative (0.0379 ns$^{-1}$) rates
with efficiency 0.66
(a), and in the
limit $n_{eh}\rightarrow \infty$ for the F\"orster (0.175 ns$^{-1}$) and
radiative (0.0379 ns$^{-1}$) rates with efficiency 0.82 (b).
}
\label{Fig3}
\end{figure}

In Fig. \ref{Fig3}, we show  the F\" orster and the radiative
transition rates,  in the classical/intermediate regime as obtained from formulas
(\ref{Forster classical rate}) and
(\ref{radiative classical rate}),
respectively, for $n_{eh}=10^{12} \text{ cm}^{-2}$ as a function of temperature
(a) and for $T=300$ K
as a function of $n_{eh}$ (b).
The dashed-dotted line is the efficiency of the F\"orster transfer.
The straight-line segments
indicate the asymptotic degenerate values given by formulas
(\ref{Forster degenerate rate})
and
(\ref{radiative degenerate rate}) in the limit $T\rightarrow 0$ (a) (0.0749
ns$^{-1}$ for the F\"orster rate and
0.0379 ns$^{-1}$ for the radiative rate with efficiency 0.66),
and in the limit $n_{eh}\rightarrow \infty$ (b) (0.175 ns$^{-1}$ for the F\"orster
rate and  0.0379 ns$^{-1}$ for the radiative rate with efficiency 0.82).
Independent of
temperature and carrier density,
the energy transfer rate is always higher than the radiative transition rate
in the QW.
In the classical regime both rates scale approximately as $1/T$ and
are linear in $n_{eh}$,
whereas in the degenerate case $1/\tau _F$ and $1/\tau _r$ are
only weakly dependent on temperature and carrier density.
The crossover
between the degenerate and the intermediate  regime is very pronounced
and occurs at 34.7 K (a) and
$8.6\times 10^{12}\text{ cm}^{-2}$ (b).
The crossover between the
intermediate and classical regime is at temperature $2\pi n_{eh}/2m_e$=139 K (a)
and at density $2m_eT/2\pi=2.2\times 10^{12}\text{ cm}^{-2}$ (b), and
is hardly noticeable. The efficiency is high and nearly constant
across the different regimes, so
the optimal regime is at the maximum of the ET rate, that
is, around the
hole degeneracy point that occurs at $n_{eh}=2m_h T/2\pi$.

\section{Energy transfer into discrete QD states}
\label{Discrete QD states}

Throughout the paper, we assumed that the QD density of states was
constant on the energy scale corresponding to
variations of the transition
matrix element of the free QW excitons. Experimentally, this
situation occurs if the band gap
of the QDs is much smaller than the band gap of the QW,
so that the final states participating in the F\"orster transfer are in the
quasicontinuum of high-energy QD excitations.
If the band gaps of the QDs and the QW are close to
each other, then the transfer  excites discrete low-energy QD
excitons. For the case of dispersionless localized QW excitons this will only change numerical prefactors in the F\"orster
transition rate. However, as we show below, in
the case of free QW
excitons or free carriers, we obtain a new behavior for the ET rate if
the width of
the QD transition is small compared to the scale, on which significant
variations of the ET transition matrix element occur.

If the width
of the QD exciton is smaller than $1/2MR^2$, we can approximate
the line by a delta function and obtain
\be
{1\over \tau _{F,\delta }(K)}\sim
\left({e^2 \over \epsilon}\right)^2
|d_D|^2 |d_W|^2
{n_D \over a_B^2} K^2 e^{-2KR}
\delta \left( {K^2 \over 2M}-E_D\right),
\ee
where $E_D$ is the energy of the QD exciton measured with respect to
the bottom of the QW exciton band. The use of
the Fermi Golden Rule is justified if the rate $1/{\tau _{F,\delta }(K)}$ is smaller than the 
dephasing rate of the QD exciton (determined by its linewidth).
The latter condition is usually satisfied in real QW/nanocrystal QD systems, 
because the ET rate is in the sub-$\mu$eV range, while the widths of 
nanocrystal transitions even at low temperatures are in the sub-meV-to-meV
range \cite{Empedocles,Han}.

The thermal averaging will lead to the replacement
\be
{N_D \over R^2}(E_G)f(2MTR^2)
\rightarrow
{M E_D \over T} e^{-2R \sqrt{2ME_D}-{E_D\over T}}.
\ee
We see that at large $T$, the dependence of the rate is still
$\propto 1/T$ as in the case of high-energy
quasicontinuous QD excitations. However, the QW-QD distance dependence changes
completely. While the ET rates calculated in the
Sections \ref{Excitons in the QW} and \ref{Free carriers} follow 
approximately the  $R^{-4}$ dependence
[$f(2MTR^2)\propto R^{-2}$ for large $R$], the ET rates
in the case of transfer into discrete QD states  depends 
 exponentially on the QW-QD separation. The rate of the exponential
distance decay can be controlled by the QW-QD energy offset   $E_D$.
In the resonant case, $E_D$ is non zero but is determined by the inhomogeneous
broadening of the QD transition energy. Typical QDs are synthesized
with a narrow size dispersion of approximately 7 \%, which translates into an 
energy variation $\Delta E$ of
 $\sim$50 $meV$ around a center energy of 2.2 eV. Assuming that 
in the ``quasi-resonant" case $E_D=\Delta E$, we find 
that the ET rate
decreases on a very short length scale of $\sim$5 \AA.

\section{Experiment: Energy transfer between an InGaN QW and CdSe QDs}
\label{Experiment}

In this section, we compare our theoretical calculations with results of the measurements
that we perform for the hybrid structure schematically shown in Fig.\ \ref{fig with geometry}. 
In our experimental work we use optical excitation (see below), therefore, 
the QW in our device does not have metal contacts.
The structure consists of a close-packed monolayer of CdSe/ZnS core/shell 
nanocrystals deposited by the Langmuir-Blodgett technique on an InGaN/GaN heterostructure.
The $30$ \AA\ wide InGaN quantum well is capped with a thin, $30$ \AA\ thickness GaN top barrier layer. 
We excite the hybrid QW/QD structure at 266 nm with 100 femtosecond laser pulses from 
a frequency-tripled, amplified Ti:sapphire laser and measure the dynamics
of the QW emission at 400 nm using a time-correlated single photon counting 
system.\cite{Marc}
First, we measure the photoluminescence (PL) dynamics from an isolated QW 
without  QDs in its
proximity. From the quadratic carrier density dependence of the PL amplitude at
zero time delay we conclude that the QW excitation can be described by non-degenerate 
free carriers. Despite the large ``nominal" exciton binding energy (675 K), charge carriers
in InGaN QWs do not form excitons at room temperature because of strong piezo-electric
fields \cite{PEFref}. 

\begin{figure}
\epsfxsize=3 in
\epsfbox{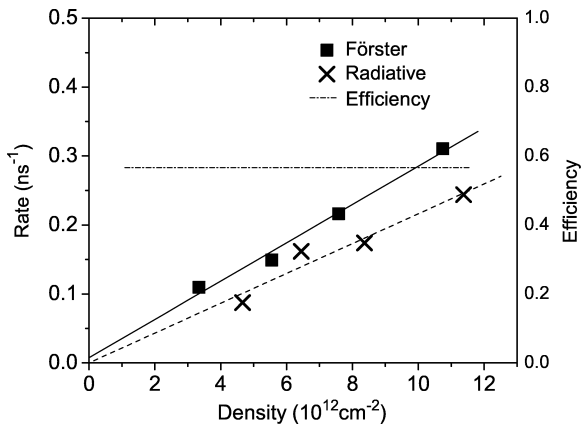}
\caption{Experimental results: F\"orster (squares) and radiative recombination rates (crosses)
as a function of excited carrier density (the solid and dashed lines are guides to the eye). 
The dash-dotted line is  the ET efficiency ($\eta$ = 57 \%).}
\label{Fig4}
\end{figure}

For the free carrier case we would expect a PL decay rate that depends linearly 
on the carrier density [eq. (\ref{radiative classical rate})]. However, we find a non-vanishing decay rate of
1 ns$^{-1}$  at low carrier 
densities that we attribute to non-radiative recombination as a result of carrier trapping at defects 
(typically observed for InGaN QWs at room temperature). 
After subtracting this
non-radiative contribution from the decay dynamics,
we obtain the radiative recombination rate (Fig. \ref{Fig4})
that shows the expected linear carrier density dependence. 
The experimental and calculated radiative rates agree within
a factor of two. 
Next, we compare the QW PL decay dynamics of the
isolated QW (decay rate $1/\tau_{w/o\ QDs}$) with the dynamics measured for the hybrid QW/QD
structure (decay rate $1/\tau_{with\ QDs}$). We find an accelarated PL decay in the hybrid structure as 
a result of ET from
the QW to the QDs. The ET rate ($1/\tau_{F}=1/\tau_{w/o\ QDs}-1/\tau_{with\ QDs}$)
is plotted in Fig. \ref{Fig4}. Both the linear carrier density
dependence and the absolute values of the measured ET rate are in agreement with eq. (\ref{Forster classical rate})
and Fig. \ref{Fig3}, respectively. The ET efficiency of 57 \%
\footnote{To compare measured and calculated ET efficiencies we
only consider radiative recombination and disregard non-radiative decay processes.} that we calculated 
from the measured ET and radiative recombination rate is close to the theoretical value of 80 \%.

\section{Summary and conclusions}
\label{Summary and conclusion}
We have studied the rate of  non-radiative F\"orster energy
transfer between  a quantum well and a proximal layer of nanocrystal QDs. We considered both the
low-density/low-temperature regime, in which the excitations are bound
excitons either
free or localized
[eq. (\ref{combined Forster rate for excitons}) and
(\ref{combined recombination rate for excitons})]
and the high-density/high-temperature regime in which the electrons
and holes form a plasma of free charge carriers, in the
non-degenerate
[eq. (\ref{Forster classical rate})
and (\ref{radiative classical rate})], degenerate
[eq. (\ref{Forster degenerate rate})
and (\ref{radiative degenerate rate})], or intermediate degenerate/non-degenerate
[eq. (\ref{mixed condition})] regimes.
For the numerical estimations, we used physical
parameters  from the experiments reported in Ref.\ 14.
For the case of QW excitons, we
find that the energy transfer into the QDs is optimal
if the contribution from the localized excitons is negligible and if
the temperature is on the order of the exciton kinetic energy with
momentum equal to the inverse of the distance between the quantum well
and the layer of the quantum dots.
For the case of free carriers in the quantum well,
we find that the energy transfer is optimal around the hole
degeneracy temperature. In addition, we considered ET to 
discrete QD states.
For this configuration, we found that the ET rates decay exponentially with
increasing QW-QD separation. The characteristic distance turned out 
to be very small, on
the order of a few \AA. Finally, we validated the theoretical
model by comparing our calculations with experimental results 
obtained for a hybrid InGaN QW/CdSe QD device. 
We measured  ET rates,
radiative recombination rates, and ET efficiencies and found a good agreement
with our theoretical predictions for the case of free carriers in the QW. Independently 
we confirm that under our experimental conditions  QW excitations 
are indeed present in the form of free carriers.
In conclusion, our results indicate that with a
careful design of the system (geometrical and electronic parameters),  the
F\"orster transfer can be used as an efficient ``non-contact" pumping mechanism of
nanocrystal QD-based light emitting devices.

\section{Acknowledgement} This project  was supported by
Los Alamos LDRD Funds and the Chemical Sciences, Biosciences, and
Geosciences Division of the Office of Basic Energy Sciences,
Office of Science, U.S. Department of Energy.

\bibliographystyle{apsrev}

\begin{thebibliography}{99}
\bibitem{footnote1}
We assume charge neutrality, that is $n_e=n_h\equiv n_{eh}$.
\end{thebibliography}

\end{document}